\documentclass[conference]{IEEEtran}

\usepackage{amsmath,amssymb,amsfonts}
\usepackage{graphicx}
\usepackage{url}
\usepackage{booktabs}
\usepackage{tikz}
\usepackage{subcaption}
\usepackage{enumitem}
\usepackage{mathtools}
\usepackage{algorithm}
\usepackage{algpseudocode}
\usepackage{bm}
\usepackage{comment}
\usepackage{cite}
\usepackage{threeparttable}

\definecolor{boristext}{rgb}{0.22, 0.44, 0.33}
\definecolor{boriscomments}{rgb}{0.83, 0.0, 0.0}
\definecolor{davidcomments}{rgb}{0.0, 0.0, 0.83}
\definecolor{miguelcomments}{rgb}{0.5, 0, 0.8}

\title{\textsc{BRAVR}: An AP-Assisted Online DRL Mechanism for Interactive VR Bitrate Adaptation over Wi-Fi}

\author{\IEEEauthorblockN{Miguel Casasnovas, Francesc Wilhelmi, and Boris Bellalta}
\IEEEauthorblockA{\textit{Department of Engineering} \\
\textit{Universitat Pompeu Fabra}\\
Barcelona, Spain \\
name.lastname@upf.edu
}
}

\IEEEoverridecommandlockouts

\begin{document}

\IEEEpubid{\makebox[\columnwidth]{© 2026 IFIP\hfill}
\hspace{\columnsep}\makebox[\columnwidth]{ }}

\maketitle

\begin{abstract}
Interactive virtual reality~(VR) streaming over Wi-Fi requires stringent latency and reliability guarantees, which become increasingly difficult to achieve under dynamic channel conditions and shared-medium contention. These challenges make real-time bitrate adaptation a critical yet fundamentally difficult control problem, particularly under limited visibility of network conditions. This paper formulates VR bitrate adaptation as a network-aware, online decision-making problem and proposes \textsc{BRAVR}, a decentralized deep reinforcement learning~(DRL) algorithm that combines application-layer observations with lightweight wireless statistics collected at the Wi-Fi access point~(AP) to balance visual quality, streaming performance, and multi-user airtime fairness. We implement \textsc{BRAVR} in a real VR streaming system and evaluate it on a physical Wi-Fi testbed against the VR-tailored \textsc{NeSt-VR} algorithm and an ablated \textsc{BRAVR} variant without AP telemetry. Experimental results show that, in the evaluated scenarios, \textsc{BRAVR} consistently achieves its design objectives, delivering robust quality of service~(QoS) while avoiding sustained airtime overutilization. It also outperforms the AP-blind ablation, demonstrating the value of incorporating network-level information into the bitrate adaptation control loop. Overall, these results demonstrate the potential of AP-assisted online learning for decentralized interactive VR streaming over commodity Wi-Fi and provide practical insights into bitrate adaptation in shared wireless environments.
\end{abstract}


\section{Introduction} 
Virtual reality~(VR) streaming offloads rendering and encoding to edge or cloud servers, enabling high-fidelity applications on lightweight head-mounted displays~(HMDs) while reducing on-device computation and energy consumption. To support user mobility and untethered operation, reliable wireless connectivity is therefore essential. Among available wireless technologies, Wi-Fi is a natural candidate because of its ubiquity and access to wide unlicensed bands, which enable high data rates. In principle, Wi-Fi can meet the stringent quality of service~(QoS) requirements of VR, including high frame rates, low latency, and near-lossless delivery~\cite{hossain2023survey}. However, as a contention-based shared medium, Wi-Fi is susceptible to interference, congestion, and time-varying channel conditions---factors that often introduce delay, jitter, and packet loss, ultimately degrading users' quality of experience~(QoE).

Under this setup, video bitrate selection becomes a critical control parameter. High bitrates improve visual quality but increase network load, potentially causing queueing delays and loss, whereas conservative bitrates may underuse capacity---a trade-off balanced by adaptive bitrate~(ABR) algorithms. Classical ABR relies on hand-crafted heuristics such as throughput estimation~\cite{yin2015control} and buffer occupancy rules~\cite{spiteri2020bola}. Although efficient and widely adopted, these methods use simplified assumptions that limit adaptation to complex or unseen conditions. To address these limitations, machine learning~(ML), particularly reinforcement learning~(RL), has increasingly been explored for bitrate adaptation~\cite{mao2017neural, hafez2023reinforcement, naresh2023ppo}. Such approaches can learn adaptive policies through interaction, making them better suited to complex and non-stationary conditions.

In parallel, edge computing has emerged as a promising paradigm for improving adaptive bitrate decisions. Edge-assisted adaptation mechanisms have been studied in standardization (e.g., MPEG-DASH SAND~\cite{ISO23009_5_2017}) and research~\cite{mehrabi2018edge, kleinrouweler2016delivering, wu2024ap}. By exposing network-side measurements unavailable to clients, edge-assisted mechanisms provide additional network context that can inform bitrate decisions, including those made by ML-based adaptation algorithms. For example, Wi-DASH~\cite{wu2024ap} applies deep reinforcement learning~(DRL) to HTTP-based bitrate adaptation using physical~(PHY)- and medium access control~(MAC)-layer statistics from an OpenWrt Wi-Fi access point~(AP).

Despite these advances, most ABR algorithms target conventional video-on-demand, whose underlying assumptions---particularly regarding buffering and latency tolerance---do not hold for \emph{interactive} VR streaming. This has motivated dedicated methods such as \textsc{EVeREst}~\cite{liubogoshchev2021everest, korneev2024model} and \textsc{NeSt-VR}~\cite{maura2024experimenting, casasnovas2025nest}. However, existing solutions remain predominantly heuristic and rely solely on application-layer measurements.
To date, learning-based and network-assisted bitrate adaptation remains largely unexplored for interactive VR streaming. Existing work instead applies learning to radio-resource allocation~\cite{kougioumtzidis2025deep} and system-parameter tuning~\cite{sun2022enabling}, and uses edge assistance for computation offloading~\cite{singh2022mobility} and scheduling~\cite{lu2024practical}. Bitrate adaptation, in contrast, is more common in less latency-sensitive panoramic (360$^\circ$) streaming, which typically uses multi-second buffers~\cite{jiang2020reinforcement, kan2021rapt360, quan2020reinforcement, li2022federated}.

This paper formulates bitrate adaptation as an \emph{online}, non-episodic learning problem and presents {\textsc{\textbf{BRAVR}}}, a decentralized AP-assisted \textbf{B}it\textbf{R}ate \textbf{A}daptation algorithm for interactive \textbf{VR} streaming over Wi-Fi based on DRL. It targets streaming quality while discouraging airtime overutilization in multi-user scenarios, and is integrated into the open-source Air Light VR~(ALVR)~\cite{alvr} platform. We evaluate the proposed AP-assisted policy on a physical testbed against \textsc{NeSt-VR} and an ablated \textsc{BRAVR} variant without AP telemetry.

The main contributions of this paper are:
\begin{itemize}
\item Integration of lightweight Wi-Fi telemetry from OpenWrt-compatible APs into the bitrate control loop.
\item Design of a decentralized DRL mechanism that combines application- and network-level information to improve visual quality while satisfying VR QoS constraints and promoting multi-user airtime fairness.
\item Evaluation on a physical testbed demonstrating the effectiveness of online learning and the benefits of AP-level information for bitrate control.
\end{itemize}

To the best of our knowledge, this is the first end-to-end integration of online DRL and network-assisted adaptation for interactive VR over commodity Wi-Fi. As an experimental proof of concept, the study has practical constraints that limit exhaustive scalability and hyperparameter-sensitivity analyses.


\section{System Architecture}
\label{sec:system}

\subsection{System Overview}
We consider a Wi-Fi basic service set~(BSS) supporting one or more asynchronous and independently controlled interactive VR sessions. Each session pairs a remote \emph{streaming server} that renders and encodes content with an associated \emph{client} (an HMD) that receives and displays it.
Communication is bidirectional: HMDs send tracking, control, and telemetry data in the uplink, while servers send encoded video frames, audio, control information, and haptic feedback in the downlink. Both directions are sensitive to delay and loss, and impairments in either direction can degrade user experience. For instance, uplink impairments can affect pose estimation and interaction responsiveness, whereas downlink impairments can reduce visual continuity and cause perceptible artifacts.
Among all traffic flows, video dominates network load, and its encoding bitrate---whether fixed or adaptive---largely determines channel occupancy and airtime consumption. Thus, each bitrate decision affects contention and resource availability in the shared medium, coupling concurrent sessions. Bitrate selection therefore emerges as a central control variable for managing resource usage and overall system performance.

\subsection{Learning-based Policy for Bitrate Adaptation}

Each server runs a \emph{decentralized} learning-based policy without centralized coordination or explicit signaling. At each decision epoch, it selects the next encoding bitrate based on heterogeneous \emph{application-level information} (e.g., client feedback and server statistics) and, when available from an external source (e.g., the AP), \emph{wireless network information} about observed channel and medium conditions. Sessions therefore adapt independently to local conditions while incorporating shared-medium observations.

\subsection{AP-Assisted Network Observability}

\begin{figure}[t!]
    \centering
    \includegraphics[width=0.9\linewidth] {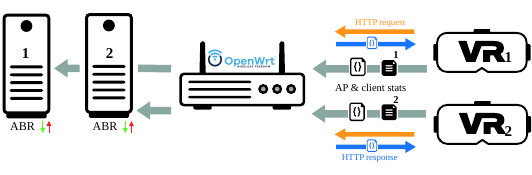}
    \caption{\textbf{AP-assisted VR streaming architecture.}
    }
    
    \label{fig:system}
\end{figure}

To provide visibility into wireless conditions, the Wi-Fi AP serves as a shared observation point. As the downlink's final hop, it has access to network and channel information unavailable to HMDs, which it can expose to streaming entities through, for instance, a lightweight, on-demand pull interface.
In our endpoint-driven implementation, HMDs periodically retrieve AP telemetry and forward it to their servers, which incorporate this information into bitrate adaptation alongside server-side measurements and client feedback (Fig.~\ref{fig:system}).

\section{BRAVR: Learning Formulation}\label{sec:learning_based}

We formulate bitrate adaptation as a sequential decision problem under uncertainty and detail \textsc{BRAVR} below.

\subsection{Problem Formulation}\label{sec:problem}

We formulate the problem as an infinite-horizon Markov decision process~(MDP)~\cite{sutton1998reinforcement}, $\mathcal{M} = (\mathcal{S}, \mathcal{A}, P, r, \gamma)$, where $\mathcal{S}$ denotes the state space, $\mathcal{A}$ the action space of size $K$, $P$ the unknown state-transition dynamics, $r$ the reward function, and $\gamma \in (0,1)$ the discount factor.
At each decision epoch~$t$, the agent observes the state $s_t \in \mathcal{S}$ and selects an action $a_t \in \mathcal{A}$ according to the policy $\pi$. The environment then transitions to $s_{t+1}$ and provides reward $r_t = r(s_t,a_t)$. The objective is to learn a policy that maximizes the expected discounted return
$\mathbb{E}_{\pi}[\sum_{t=0}^{\infty}\gamma^t r_t]$.

\subsection{Observation Space}\label{sec:state}

In practice, because the full system state is not directly observable, at each decision epoch~$t$, the agent instead observes a partial feature vector
\begin{equation}
o_t = \big[\, o_t^{\text{app}},\, o_t^{\text{hist}},\, o_t^{\text{net}} \,\big],
\end{equation}
where $o_t^{\text{app}}$, $o_t^{\text{hist}}$, and $o_t^{\text{net}}$ are application-, history-, and network-level features computed over the preceding interval and normalized for numerical stability.

\emph{Application-level features} $o_t^{\text{app}}$ capture QoS and streaming performance: the \emph{signed deviations from target values} of the \emph{network frame success ratio} (NFR), defined as the fraction of transmitted frames successfully received, and the \emph{video frame round-trip time} (VF-RTT), defined as the delay between frame transmission and reception of the corresponding receiver feedback; their \emph{short-term temporal variations}, computed as first-order differences between observations; a \emph{latency spike measure}, defined as the peak-to-mean VF-RTT deviation; the \emph{bitrate efficiency}, defined as the ratio of delivered bitrate to target bitrate; and the \emph{current bitrate}.

\emph{History-level features} $o_t^{\text{hist}}$ encode recent control decisions: the \emph{holding duration}, defined as the number of decision steps since the last bitrate change; and the \emph{previous action}.

\emph{Network-level features} $o_t^{\text{net}}$, derived from AP telemetry, capture stream-specific and shared-medium wireless conditions: the stream's \emph{downlink modulation and coding scheme} (MCS), indicating the achievable PHY rate; the stream \emph{packet retransmission rate}; the network-wide \emph{channel utilization}, defined as the fraction of time the medium is busy; the \emph{inverse active VR-client count}, defined as the reciprocal of the number of active VR clients in the BSS; the stream-specific \emph{airtime fraction}, representing the proportion of channel occupancy attributed to the stream; and the network-wide \emph{airtime fairness} across active VR streams, measured using Jain's fairness index.

\subsection{Action Space}\label{sec:action}

At each decision epoch~$t$, the agent selects $a_t \in \mathcal{A}$ based on the observation $o_t$:
\begin{equation}
\mathcal{A} = \{-1, 0, +1\}.
\end{equation}
The actions \emph{decrease}, \emph{maintain}, or \emph{increase} the encoding bitrate, respectively, using a predefined ladder. Compared with selecting absolute bitrate levels, adjacent-level transitions reduce the action space and enable gradual adaptation, avoiding abrupt bitrate changes that could degrade QoE.


\subsection{Action Shielding}

\begin{figure}[t!]
    \centering
    \begin{subfigure}[t]{\linewidth}
        \centering
        \includegraphics[width=0.85\linewidth]{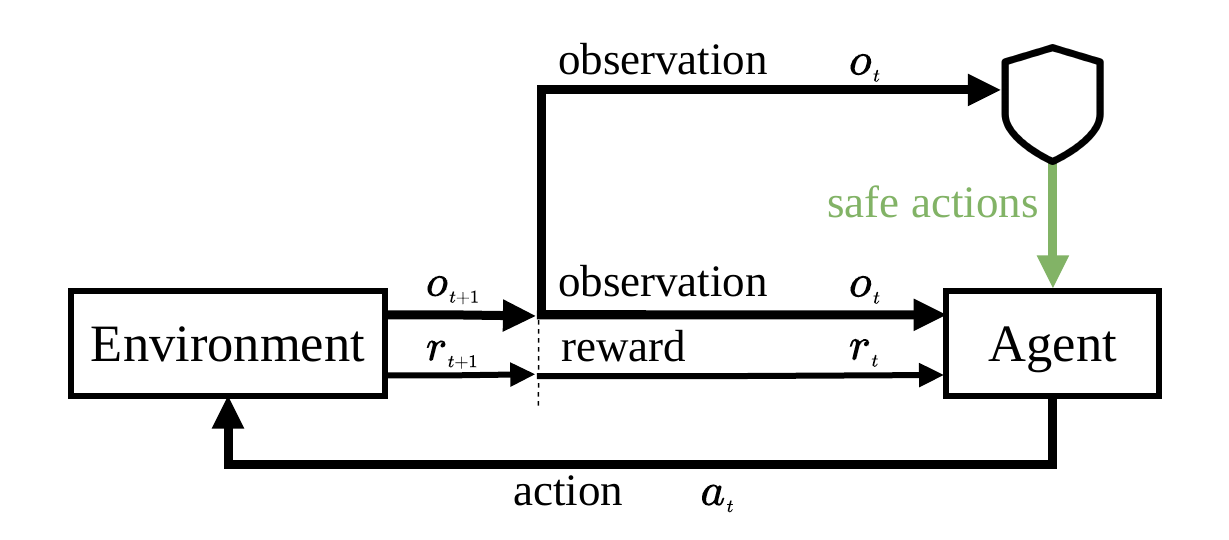}
    \end{subfigure}
    \caption{\textbf{Safe reinforcement learning with preemptive action shielding.} The agent--environment interaction loop.}
    \label{fig:shielded_rl}
\end{figure}

To support safe online operation, we employ \emph{preemptive action shielding}~\cite{alshiekh2018safe} (Fig.~\ref{fig:shielded_rl}). At each decision epoch~$t$, the action set is conditioned on the current observation: bitrate increases are forbidden under unfavorable conditions (i.e., when QoS metrics deviate beyond predefined tolerance thresholds), and invalid actions at the bitrate ladder boundaries are masked. The agent thus selects from a filtered set $\tilde{\mathcal{A}}(o_t)\subseteq\mathcal{A}$, excluding unsafe or ineffective actions. This runtime safety layer operates independently of policy learning, helping prevent unsafe actions that may cause instability or disconnections in interactive, real-time applications.

\subsection{Reward Function}\label{sec:reward}

At each decision epoch~$t$, the agent receives a scalar reward computed from the current observation and selected action:
\begin{equation}\label{eq:reward}
r_t = w_b\, b_t - \sum_{k \in \mathcal{K}} w_k\, p_t^{k}, \quad
{\scriptstyle \mathcal{K} = \{\mathrm{rtt}, \mathrm{nfr},  \mathrm{beh}, \mathrm{air}\}},
\end{equation}
where $b_t$ denotes the \emph{bitrate}, $p_t^{k}$ are penalty terms (clipped for stability), and $w_{b}$ and $w_{k}$ are weights.

The \emph{bitrate} $b_t$, normalized over the bitrate ladder, encourages higher visual quality. A linear min--max normalization of bitrate is adopted for simplicity and interpretability.

The \emph{video frame round-trip time penalty} $p_t^{\mathrm{rtt}}$ and the \emph{network frame success ratio penalty} $p_t^{\mathrm{nfr}}$ penalize deviations from target values under predefined tolerances, encouraging low streaming latency and reliable frame delivery, respectively.

The \emph{behavioral penalty} $p_t^{\mathrm{beh}}$ acts as a soft regularizer, penalizing bitrate decreases under favorable conditions and increases under unfavorable ones, discouraging unnecessary transitions.

The \emph{airtime fairness penalty} $p_t^{\mathrm{air}}$ penalizes airtime usage above a nominal fair share, defined as the available airtime budget equally divided among active VR users, rather than enforcing equal airtime allocation. This assumes that concurrent VR streams can reach comparable demand levels, such that exceeding the nominal share indicates unfair resource usage.

This linear additive structure follows common design patterns in QoE-oriented ABR formulations but differs fundamentally from classical buffer-centric models such as~\cite{yin2015control}. Reliability and latency, major contributors to VR QoE~\cite{rossi2024qoe}, are treated as \emph{soft objectives} rather than hard constraints, avoiding constrained optimization formulations.


\subsection{Learning Algorithm: Deep $n$-step Expected SARSA}

\begin{figure}[t!]
    \centering
    \begin{subfigure}[t]{\linewidth}
        \centering
        \includegraphics[width=0.9\linewidth]{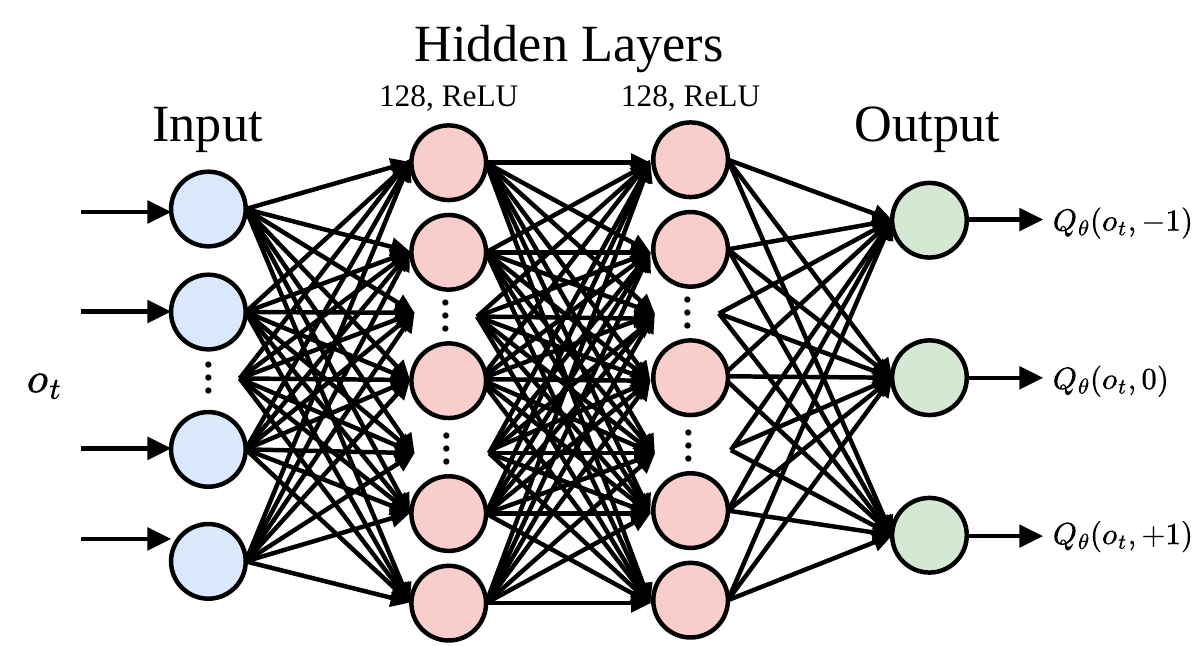}
    \end{subfigure}
    \caption{\textbf{Feedforward network for estimating $Q_\theta(o_t,a)$.} Two fully connected ReLU layers map the observation to one $Q$-value for each action.}
    \label{fig:nn}
\end{figure}

We adopt \emph{Deep $n$-step Expected SARSA}~\cite{sutton1998reinforcement, van2009theoretical}, a primarily \emph{on-policy temporal-difference~(TD)} method for online adaptation in non-stationary environments such as Wi-Fi networks, where action effects can be delayed and temporally coupled. The \emph{$n$-step} return accounts for delayed action effects by propagating rewards across multiple steps, while the \emph{expected} formulation reduces variance by marginalizing over the policy. A feedforward neural network provides \emph{deep function approximation} for the action-value function $Q_\theta(o_t,a)$ (Fig.~\ref{fig:nn}), enabling it to handle continuous observation spaces.

Learning uses TD updates with error $\delta_t = y_t - Q_\theta(o_t,a_t)$, where $y_t$ denotes the $n$-step expected SARSA target.
The target is defined as
\begin{equation}
y_t = \sum_{k=0}^{n-1} \gamma^k r_{t+k}
+ \gamma^n \sum_{a \in \tilde{\mathcal{A}}(o_{t+n})} \pi(a \mid o_{t+n}) Q_{\theta^-}(o_{t+n}, a),
\end{equation}
where $\gamma \in (0,1)$ denotes the discount factor, $n$ the horizon, $\pi$ the current policy over admissible actions, and $\theta^-$ denotes the parameters of a slowly updated target network, updated via $\theta^- \leftarrow (1-\beta)\theta^- + \beta\theta$ with Polyak coefficient $\beta$~\cite{polyak1992acceleration}. The parameters $\theta$ are optimized by minimizing the \emph{Huber loss}~\cite{huber1992robust}; the resulting gradients are clipped by their norm before applying the \emph{Adam} optimizer~\cite{kingma2014adam} with learning rate $\alpha$ to ensure stable updates.
Action selection follows a policy $\pi$ defined as a convex mixture of a \emph{Boltzmann (softmax) policy} $\pi_{\text{B}}$ with temperature~$T$~\cite{sutton1998reinforcement} and a uniform distribution $\pi_{\text{U}}$ over admissible actions:
\begin{equation}
    \pi(a \mid o_t) = (1-\varepsilon)\,\pi_{\text{B}}(a \mid o_t) + \varepsilon\,\pi_{\text{U}}(a \mid o_t),
\end{equation}
where $\varepsilon \in [0,1]$ is a fixed mixing coefficient. Both components are defined over the filtered set $\tilde{\mathcal{A}}(o_t)$, so masked actions receive zero probability.

\section{Experimental Setup and Methodology}\label{sec:evaluation}

\subsection{Experimental Setup}

Experiments use a dedicated VR-streaming BSS in a university-office Wi-Fi testbed. Single- and multi-user setups include one or two laptops as servers, one Wi-Fi AP running OpenWrt~\cite{openwrt}, and one or two Meta Quest~2 HMDs as clients. Servers connect via $1$~Gbps Ethernet; HMDs use a $40$~MHz Wi-Fi~5 (IEEE~802.11ac) channel in the $5$~GHz band with 23~dBm transmit power and no multi-user features.\footnote{Under Wi-Fi~6, the AP supported stable high-bitrate streaming, obviating adaptation and motivating the more constrained Wi-Fi~5 configuration.} Users are either near the AP (Location~A, approximately $-48$~dBm), farther away (Location~B, approximately $-82$~dBm), or moving between them in mobility scenarios. Hardware is summarized in Table~\ref{tab:equipment}. Servers and clients run a custom ALVR v20.6.0 fork~\cite{alvr_fork} integrating the learning pipeline and AP telemetry retrieval. SteamVR Home provides the VR workload. Streaming uses HEVC encoding, a configured per-eye resolution of $2144\times2336$, 90~fps, foveated encoding, and a two-frame client buffer (as recommended in~\cite{korneev2024model}).

\vspace{0.08cm}
\noindent\textit{AP-level telemetry collection and exchange:} The OpenWrt AP runs a lightweight monitoring service that collects wireless statistics using standard utilities (\textit{iw}, \textit{iwinfo}, \textit{ip}). The service provides the collected information through a stateless HTTP interface over TCP using short-lived request--response interactions, and returns a compact JSON payload containing channel-level metrics (e.g., active and busy times), per-client wireless metrics (e.g., received signal strength indicator~(RSSI), downlink/uplink transmission durations, and MCS), and metadata identifying active VR clients. 
In the considered endpoint-driven implementation, HMDs periodically issue HTTP GET requests and forward the response to their server over the streaming system's existing messaging channel (Fig.~\ref{fig:system}). The exchange incurs an overhead of 4--8~KB/s per session, negligible relative to video traffic.

\begin{table}[t]
    \centering
    \caption{\textbf{Testbed equipment.}}
    \footnotesize
    \begin{tabular}{@{}lll@{}}
\toprule
     \textbf{Laptop 1} & Model & Dell G15 5521 \\
    & GPU  & NVIDIA GeForce RTX 3060 Mobile  \\
    & CPU  & Intel Core i7-12700H \\
    \midrule
    \textbf{Laptop 2} & Model & Dell Pro Max 16 \\
    & GPU  & NVIDIA RTX Pro 1000 Blackwell  \\
    & CPU  & Intel Core Ultra 7 265H \\
    \midrule
    \textbf{AP} & Model & ASUS TUF-AX4200 \\
    & Firmware  & OpenWrt 23.05.5 \\
    \midrule
    \textbf{HMDs} & Model & Meta Quest~2 \\
     \bottomrule
    \end{tabular}
    \label{tab:equipment}
\end{table}

\begin{table}[t]
\centering
\caption{\textbf{Learning hyperparameters.}}
\label{tab:learning_config}
\begin{tabular}{ccl}
\toprule
\textbf{Symbol} & \textbf{Value} & \textbf{Meaning} \\
\midrule
$n$ & $3$ & n-step return horizon \\
$\gamma$ & $0.8$ & discount factor \\
$\alpha$ & $10^{-3}$ & learning rate \\
$\beta$ & $0.01$ & Polyak averaging coefficient \\
$T$ & $0.25$ & softmax temperature \\
$H$ & $2 \times 128$ & hidden layers (units per layer) \\
$\varepsilon_{\text{train}}$ & $0.05$ & 
mixing coefficient (training) \\
$\varepsilon_{\text{eval}}$ & $0$ & 
mixing coefficient (evaluation) \\
$\|\nabla\|_2$ & $1.0$ & gradient L2 norm clipping threshold \\
\midrule
$w_b$ & $1.0$ & bitrate reward weight \\
$w_{\mathrm{rtt}}$ & $1.5$ & latency weight \\
$w_{\mathrm{nfr}}$ & $1.5$ & reliability weight \\
$w_{\mathrm{beh}}$ & $0.2$ & behavior weight \\
$w_{\mathrm{air}}$ & $10.0$ & airtime fairness weight \\
\bottomrule
\end{tabular}
\end{table}

\begin{table*}[t]
\centering
\footnotesize
\caption{\textbf{Single-user performance under stationary and mobility conditions.} Metrics are means~$\pm$~standard deviations over evaluation sessions; cell colors indicate QoS compliance (black: within target, blue: tolerated deviation, red: beyond limits), and bold values indicate the best result.}
\label{tab:single_user_all}

\begin{subtable}[t]{0.78\textwidth}
\centering
\footnotesize
\caption{Stationary conditions.}

\begin{tabular*}{\textwidth}{@{\extracolsep{\fill}}l
c@{\,$\pm$\,}c c@{\,$\pm$\,}c
c@{\,$\pm$\,}c c@{\,$\pm$\,}c
c@{\,$\pm$\,}c c@{\,$\pm$\,}c@{}}
\toprule
& \multicolumn{4}{c}{\textbf{NeSt-VR}}
& \multicolumn{4}{c}{\textbf{\textsc{BRAVR}$^{-}$}}
& \multicolumn{4}{c}{\textbf{\textsc{BRAVR}$^{+}$}} \\

& \multicolumn{2}{c}{\textbf{A}} & \multicolumn{2}{c}{\textbf{B}}
& \multicolumn{2}{c}{\textbf{A}} & \multicolumn{2}{c}{\textbf{B}}
& \multicolumn{2}{c}{\textbf{A}} & \multicolumn{2}{c}{\textbf{B}} \\
\midrule

Target bitrate
& $115.3$ & $4.3$ & $97.5$ & $28.6$
& $108.2$ & $2.5$ & $69.9$ & $8.8$
& $\bm{122.6}$ & $\bm{3.5}$ & $\bm{110.1}$ & $\bm{11.1}$ \\
\midrule

NFR
& $\bm{100.0}$ & $\bm{0.0}$ & \textcolor{blue!80!black}{$97.6$} & \textcolor{blue!80!black}{$3.2$}
& $99.9$ & $0.1$ & \textcolor{blue!80!black}{$\bm{98.6}$} & \textcolor{blue!80!black}{$\bm{1.9}$}
& $\bm{100.0}$ & $\bm{0.0}$ & \textcolor{blue!80!black}{$97.6$} & \textcolor{blue!80!black}{$1.5$} \\

VF-RTT
& $\bm{11.7}$ & $\bm{0.9}$ & $\bm{19.1}$ & $\bm{4.7}$
& $12.8$ & $1.1$ & $21.1$ & $1.7$
& $12.4$ & $0.6$ & $21.3$ & $2.2$ \\
\midrule

Switch rate
& $\bm{18.4}$ & $\bm{2.3}$ & $\bm{39.1}$ & $\bm{9.8}$
& $58.9$ & $5.2$ & $63.8$ & $2.2$
& $52.5$ & $11.8$ & $67.9$ & $8.5$ \\
\bottomrule
\end{tabular*}
\end{subtable}

\vspace{1em}

\begin{subtable}[t]{0.78\textwidth}
\centering
\footnotesize
\caption{Mobility conditions.}

\begin{tabular*}{\textwidth}{@{\extracolsep{\fill}}l
c@{\,$\pm$\,}c c@{\,$\pm$\,}c
c@{\,$\pm$\,}c c@{\,$\pm$\,}c
c@{\,$\pm$\,}c c@{\,$\pm$\,}c@{}}
\toprule
& \multicolumn{4}{c}{\textbf{NeSt-VR}}
& \multicolumn{4}{c}{\textbf{\textsc{BRAVR}$^{-}$}}
& \multicolumn{4}{c}{\textbf{\textsc{BRAVR}$^{+}$}} \\

& \multicolumn{2}{c}{\textbf{A$\to$B}} & \multicolumn{2}{c}{\textbf{B$\to$A}}
& \multicolumn{2}{c}{\textbf{A$\to$B}} & \multicolumn{2}{c}{\textbf{B$\to$A}}
& \multicolumn{2}{c}{\textbf{A$\to$B}} & \multicolumn{2}{c}{\textbf{B$\to$A}} \\
\midrule

Target bitrate
& $107.3$ & $4.0$ & $106.3$ & $6.5$
& $108.2$ & $3.6$ & $102.4$ & $14.2$
& $\bm{122.9}$ & $\bm{2.6}$ & $\bm{115.6}$ & $\bm{6.0}$ \\
\midrule

NFR
& $\bm{99.0}$ & $\bm{0.6}$ & $\bm{99.8}$ & $\bm{0.1}$
& \textcolor{blue!80!black}{$98.8$} & \textcolor{blue!80!black}{$0.9$} & \textcolor{blue!80!black}{$95.8$} & \textcolor{blue!80!black}{$4.6$}
& \textcolor{blue!80!black}{$97.8$} & \textcolor{blue!80!black}{$1.8$} & \textcolor{blue!80!black}{$98.0$} & \textcolor{blue!80!black}{$2.2$} \\

VF-RTT
& $\bm{18.2}$ & $\bm{2.7}$ & $\bm{15.3}$ & $\bm{0.4}$
& $21.6$ & $3.2$ & $18.9$ & $5.2$
& $19.5$ & $1.4$ & $18.1$ & $2.3$ \\
\midrule

Switch rate
& $\bm{48.7}$ & $\bm{8.2}$ & $\bm{45.7}$ & $\bm{8.1}$
& $64.3$ & $4.9$ & $65.2$ & $12.0$
& $78.2$ & $8.3$ & $70.7$ & $14.1$ \\
\bottomrule
\end{tabular*}

\begin{tablenotes}[flushleft]
\footnotesize
\item Target bitrate [Mbps], network frame success ratio~(NFR) [\%], video frame round-trip time~(VF-RTT)~[ms], and bitrate switching rate [changes/min].
\end{tablenotes}

\end{subtable}

\end{table*}

\subsection{Experimental Methodology}\label{sec:exp_meth}

\textsc{BRAVR} (henceforth \textsc{BRAVR}$^{+}$) is evaluated against \textsc{BRAVR}$^{-}$, an ablated variant without AP telemetry. In single-user scenarios, the airtime penalty is inactive, so the ablation isolates the effect of AP observations. In multi-user scenarios, the penalty requires AP telemetry and is omitted from \textsc{BRAVR}$^{-}$, meaning the ablation also changes the reward formulation. The evaluation also includes \textsc{NeSt-VR}~\cite{maura2024experimenting, casasnovas2025nest}, a heuristic VR ABR method that adjusts bitrate in discrete steps using probabilistic threshold rules based on latency and reliability signals. \textsc{NeSt-VR} serves as a strong non-learning baseline, having demonstrated robust performance across single- and multi-user scenarios and outperforming constant-bitrate~(CBR) strategies, general-purpose state-of-the-art controllers such as Google Congestion Control~(GCC)~\cite{carlucci2016analysis} and Network-Assisted Dynamic Adaptation~(NADA)~\cite{zhu2020network}, and the VR-tailored \textsc{EVeREst}. CBR is additionally included in multi-user tests as a non-adaptive reference.

All adaptive methods use the same 10--150~Mbps ladder in 10~Mbps steps, initialized at 10~Mbps. Bitrate decisions occur every $0.5$~s and AP telemetry is collected every $0.25$~s; \textsc{NeSt-VR} uses its \emph{Balanced} profile for symmetric changes. For both learning methods, QoS targets are $22$~ms~(VF-RTT) and $0.99$~(NFR), following the recommended values in~\cite{casasnovas2025nest}; tolerances are $22$~ms and $0.05$, setting penalty scaling and unit-penalty boundaries at $44$~ms and $0.94$, respectively. The same targets define \textsc{NeSt-VR} thresholds. Table~\ref{tab:learning_config} lists the learning hyperparameters; the reward prioritizes latency and reliability over bitrate while promoting multi-user airtime fairness.

\vspace{0.08cm}
\noindent\textit{Training and evaluation procedure.}
Learning-based methods learn online on the physical Wi-Fi testbed, without simulation pretraining, and retain parameters across training sessions, avoiding the need for simulation-to-reality transfer and enabling continued real-world refinement. In single-user scenarios, training consists of seventeen consecutive $240$~s sessions; in multi-user scenarios, each method trains in one continuous 20-minute session per configuration. All methods, including \textsc{NeSt-VR} and CBR, are evaluated over three separate $240$~s sessions, and only results from these sessions are reported. For learning-based methods, each evaluation session starts from the same trained model checkpoint and continues learning online throughout the session, without retaining updates between evaluation sessions. The dataset is publicly available~\cite{dataset_bravr}.

\vspace{0.08cm}
\noindent \textit{Utility.}
To assess the bitrate--QoS trade-off, we define the instantaneous utility as $u_i(t) = r_i(t)\cdot s_i^{\mathrm{rtt}}(t)\cdot s_i^{\mathrm{nfr}}(t)$, where $r_i(t)$ is the achieved bitrate of user~$i$ and $s_i^{\mathrm{rtt}}(t), s_i^{\mathrm{nfr}}(t) \in [0,1]$ quantify satisfaction with the VF-RTT and NFR targets ($22$~ms and $0.99$). Each satisfaction factor equals one at its target and decreases linearly with violation, reaching zero at $44$~ms and $0.94$. The utility of each user, $U_i$, is computed as the time average of $u_i(t)$ over the evaluation period.

\vspace{0.08cm}
\noindent\textit{Airtime-aware utility.}
We define $U_i^{\mathrm{air}} = U_i \cdot S_i^{\mathrm{air}}$, where $S_i^{\mathrm{air}} \in [0,1]$ is the airtime satisfaction rate, defined as the fraction of intervals in which user~$i$ stays within its nominal fair share. This metric is considered for multi-user scenarios, assuming fair-share airtime usage is desirable.


\section{Results}\label{sec:results}

\subsection{Single-User Scenario}\label{sec:results_single_user}

\begin{figure}[t]
    \begin{subfigure}[b]{\linewidth}
        \includegraphics[width=0.9\linewidth]{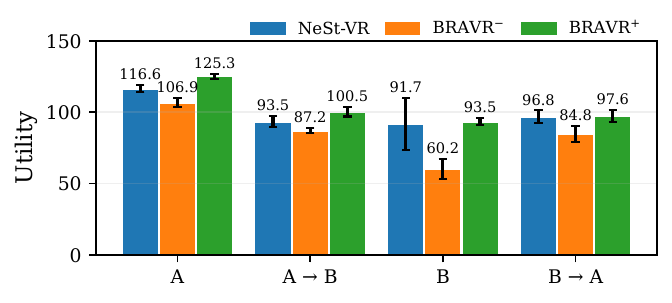}
    \end{subfigure}
    \caption{\textbf{Utility across single-user mobility stages.} Mean~$\pm$~standard deviation across evaluation sessions.}
    \label{fig:utility_su}
\end{figure}

One VR user moves between Locations~A and~B at a walking pace ($\sim1.6$~m/s), experiencing degrading and improving channel conditions due to changes in distance from the AP and the surrounding environment. The trajectory has four $60$~s stages: stationary at A, A$\to$B, stationary at B, and B$\to$A. Table~\ref{tab:single_user_all} reports stage-wise average performance; Fig.~\ref{fig:utility_su} reports stage-wise utility.

\vspace{0.08cm}
\noindent\textit{Performance.}
All approaches largely maintain mean QoS within the targets or predefined tolerance bounds. At Location~A, favorable stationary conditions allow all methods to meet the QoS targets. At Location~B, mean reliability decreases for all methods but remains within tolerance bounds. Under mobility, learning-based methods exhibit moderate reliability degradation, most notably \textsc{BRAVR}$^{-}$ during B$\to$A. This behavior is consistent with the learning methods treating QoS targets as soft constraints, allowing temporary deviations in exchange for higher long-term return, whereas threshold-based \textsc{NeSt-VR} is stricter and more conservative.
Despite this, \textsc{BRAVR}$^{+}$ achieves the highest mean selected bitrate while keeping mean QoS within the predefined tolerance bounds, resulting in the highest utility (Fig.~\ref{fig:utility_su}). Across the four single-user stages, it increases bitrate by $+21.2\%$ and QoS satisfaction (the fraction of samples meeting both strict QoS targets, capturing transient violations that temporal averaging may obscure) by $+3.8\%$ relative to \textsc{BRAVR}$^{-}$, along with a higher environment reward (median over all evaluation samples: $0.64$ vs.~$0.58$). Compared with \textsc{NeSt-VR}, which also achieves higher bitrate and tighter QoS compliance than \textsc{BRAVR}$^{-}$, \textsc{BRAVR}$^{+}$ attains a higher bitrate, whereas \textsc{NeSt-VR} adheres more tightly to the QoS targets under mobility.

\vspace{0.08cm}
\noindent\textit{Bitrate adaptation.}
Learning-based approaches exhibit more frequent adjacent-level bitrate transitions, potentially due to the stochastic nature of their policies under Boltzmann action selection and the absence of explicit switching hysteresis.
For example, \textsc{BRAVR}$^{+}$ switches at $78.2$ and $70.7$ changes/min during A$\to$B and B$\to$A, respectively, versus $48.7$ and $45.7$ for \textsc{NeSt-VR} (Table~\ref{tab:single_user_all}). Accordingly, introducing switching hysteresis could be explored to improve bitrate stability.

\vspace{0.08cm}
\noindent\textit{Policy behavior.}
An exploratory post-hoc analysis is performed for \textsc{BRAVR}$^{+}$ by applying permutation importance to a Random Forest surrogate of the learned policy during evaluation. It assesses feature relevance to bitrate-direction decisions, not causal effects on control performance. Results suggest that latency features dominate: VF-RTT deviation from target (importance 0.40), its short-term trend (0.25), and peak-to-mean VF-RTT (0.16). AP-level features also contribute: channel utilization (0.24) and stream airtime fraction (0.22). Other wireless and fairness-related features, such as MCS (0.01) and Jain's fairness index (0.00), show low individual importance.

\vspace{0.08cm}
\noindent \textit{Takeaway:}
\textsc{BRAVR}$^{+}$ uses AP telemetry to sustain higher bitrates while maintaining mean QoS within tolerance bounds under mobility. Its soft QoS constraints permit controlled deviations to improve the long-term bitrate--QoS trade-off.

\begin{table*}[t]
\centering
\footnotesize
\caption{\textbf{Multi-user performance in near--near and near--far scenarios.} Means~$\pm$~standard deviations over evaluation sessions; cell colors show QoS compliance (black: within target, blue: tolerated, red: beyond limits), bold values indicate the best result, and airtime is the percentage of channel time per user.}
\label{tab:multi_user_combined}

\begin{subtable}[t]{0.94\textwidth}
\centering
\footnotesize
\caption{Both users near the access point (U1, U2).}\label{tab:multi_user_close}

\begin{tabular*}{\textwidth}{@{\extracolsep{\fill}}l
c@{\,$\pm$\,}c c@{\,$\pm$\,}c
c@{\,$\pm$\,}c c@{\,$\pm$\,}c
c@{\,$\pm$\,}c c@{\,$\pm$\,}c
c@{\,$\pm$\,}c c@{\,$\pm$\,}c@{}}

\toprule
& \multicolumn{4}{c}{\textbf{\textsc{CBR}}}
& \multicolumn{4}{c}{\textbf{\textsc{NeSt-VR}}}
& \multicolumn{4}{c}{\textbf{\textsc{BRAVR}$^{-}$}}
& \multicolumn{4}{c}{\textbf{\textsc{BRAVR}$^{+}$}} \\

& \multicolumn{2}{c}{\textbf{U1}} & \multicolumn{2}{c}{\textbf{U2}}
& \multicolumn{2}{c}{\textbf{U1}} & \multicolumn{2}{c}{\textbf{U2}}
& \multicolumn{2}{c}{\textbf{U1}} & \multicolumn{2}{c}{\textbf{U2}}
& \multicolumn{2}{c}{\textbf{U1}} & \multicolumn{2}{c}{\textbf{U2}} \\

\midrule
Target bitrate
& $\bm{80.0}$ & $\bm{0.0}$ & $\bm{80.0}$ & $\bm{0.0}$
& ${71.2}$ & $1.6$ & ${68.3}$ & ${2.5}$
& ${77.1}$ & ${5.2}$ & $65.4$ & $6.5$
& $66.9$ & $6.8$ & $62.5$ & $4.9$ \\

\midrule
NFR
& \textcolor{red!80!black}{${88.0}$} & \textcolor{red!80!black}{$5.8$}
& \textcolor{red!80!black}{${62.4}$} & \textcolor{red!80!black}{${29.9}$}
& $99.8$ & $0.1$ & $99.5$ & $0.1$
& $99.1$ & $0.5$ & $99.2$ & $0.3$
& $\bm{100.0}$ & $\bm{0.1}$ & $\bm{100.0}$ & $\bm{0.1}$ \\

VF-RTT
& \textcolor{red!80!black}{${51.0}$} & \textcolor{red!80!black}{$13.0$}
& \textcolor{red!80!black}{${97.1}$} & \textcolor{red!80!black}{$42.6$}
& $15.7$ & $0.5$ & $16.8$ & $0.3$
& $16.9$ & $0.6$ & $16.4$ & $1.4$
& $\bm{11.2}$ & $\bm{1.9}$ & $\bm{11.3}$ & $\bm{1.8}$ \\

\midrule
Airtime
& ${46.3}$ & $0.7$ & ${47.8}$ & $0.6$
& ${45.1}$ & $0.5$ & {$44.6$} & {$1.4$}
& ${45.6}$ & $1.6$ & \bm{$44.9$ }& \bm{$2.5$}
& $\bm{44.5}$ & $\bm{2.7}$ & ${43.4}$ & ${2.6}$ \\

\midrule

Switch rate
& $\bm{0.0}$ & $\bm{0.0}$ & $\bm{0.0}$ & $\bm{0.0}$
& ${21.6}$ & ${0.7}$ & ${24.5}$ & $0.9$
& $58.7$ & $1.3$ & $61.9$ & $1.9$
& $63.2$ & $4.9$ & $57.5$ & $3.7$ \\

\bottomrule
\end{tabular*}
\end{subtable}

\vspace{1em}

\begin{subtable}[t]{0.94\textwidth}
\centering
\footnotesize
\caption{One user near the access point (U1) and one user farther from it (U2).}\label{tab:multi_user_far}

\begin{tabular*}{\textwidth}{@{\extracolsep{\fill}}l
c@{\,$\pm$\,}c c@{\,$\pm$\,}c
c@{\,$\pm$\,}c c@{\,$\pm$\,}c
c@{\,$\pm$\,}c c@{\,$\pm$\,}c
c@{\,$\pm$\,}c c@{\,$\pm$\,}c@{}}

\toprule
& \multicolumn{4}{c}{\textbf{\textsc{CBR}}}
& \multicolumn{4}{c}{\textbf{\textsc{NeSt-VR}}}
& \multicolumn{4}{c}{\textbf{\textsc{BRAVR}$^{-}$}}
& \multicolumn{4}{c}{\textbf{\textsc{BRAVR}$^{+}$}} \\

& \multicolumn{2}{c}{\textbf{U1}} & \multicolumn{2}{c}{\textbf{U2}}
& \multicolumn{2}{c}{\textbf{U1}} & \multicolumn{2}{c}{\textbf{U2}}
& \multicolumn{2}{c}{\textbf{U1}} & \multicolumn{2}{c}{\textbf{U2}}
& \multicolumn{2}{c}{\textbf{U1}} & \multicolumn{2}{c}{\textbf{U2}} \\

\midrule
Target bitrate
& $70.0$ & $0.0$ & $\bm{70.0}$ & $\bm{0.0}$
& $\bm{92.9}$ & $\bm{17.2}$ & $28.0$ & $3.0$
& $89.1$ & $6.9$ & $27.0$ & $7.5$
& ${55.4}$ & ${1.7}$ & $22.2$ & $2.5$ \\

\midrule
NFR
& $99.6$ & $0.0$ & \textcolor{red!80!black}{$88.5$} & \textcolor{red!80!black}{$0.8$}
& $99.2$ & $0.0$ & $\bm{99.8}$ & $\bm{0.2}$
& \textcolor{blue!80!black}{$98.0$} & \textcolor{blue!80!black}{$0.8$} & \textcolor{blue!80!black}{$98.8$} & \textcolor{blue!80!black}{$0.6$}
& $\bm{100.0}$ & $\bm{0.0}$ & $99.3$ & $0.9$ \\

VF-RTT
& $19.8$ & $0.2$ & \textcolor{red!80!black}{$47.0$} & \textcolor{red!80!black}{$0.8$}
& $15.6$ & $0.9$ & $15.3$ & $0.4$
& $17.3$ & $0.9$ & \textcolor{blue!80!black}{$26.5$} & \textcolor{blue!80!black}{$5.2$}
& $\bm{7.9}$ & $\bm{0.8}$ & $\bm{13.0}$ & $\bm{6.7}$ \\

\midrule
Airtime
& $41.8$ & $0.2$ & \textcolor{red!80!black}{$51.8$} & \textcolor{red!80!black}{$0.1$}
& \textcolor{red!80!black}{$58.4$} & \textcolor{red!80!black}{$8.2$} & $29.7$ & $5.6$
& \textcolor{red!80!black}{$57.0$} & \textcolor{red!80!black}{$1.36$} & $30.9$ & $3.4$
& $\bm{44.1}$ & $\bm{1.3}$ & $\bm{31.7}$ & $\bm{0.6}$ \\

\midrule

Switch rate
& $\bm{0.0}$ & $\bm{0.0}$ & $\bm{0.0}$ & $\bm{0.0}$
& $25.1$ & $1.4$ & $20.9$ & $3.4$
& $57.5$ & $6.6$ & $52.1$ & $13.0$
& $59.5$ & $4.0$ & $51.4$ & $3.4$ \\

\bottomrule
\end{tabular*}
\begin{tablenotes}[flushleft]
\footnotesize
\item Target bitrate [Mbps], network frame success ratio~(NFR) [\%], video frame round-trip time~(VF-RTT)~[ms], and bitrate switching rate [changes/min].
\end{tablenotes}

\end{subtable}

\end{table*}

\begin{figure}[t]
    \begin{subfigure}[b]{\linewidth}
        \includegraphics[width=0.9\linewidth]{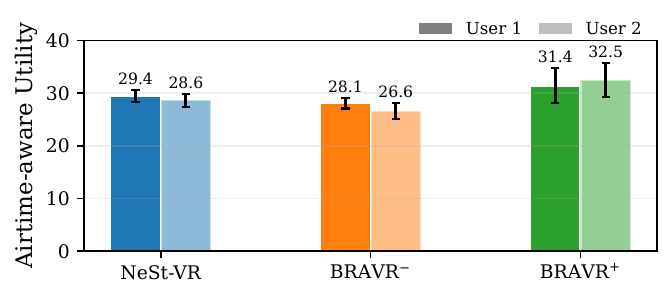}
    \end{subfigure}
    \caption{\textbf{Per-user airtime-aware utility (near--near scenario).} Mean $\pm$ standard deviation across evaluation sessions.}
    \label{fig:utility_mu_close}
\end{figure}

\begin{figure}[t]
    \begin{subfigure}[b]{\linewidth}
        \includegraphics[width=0.9\linewidth]{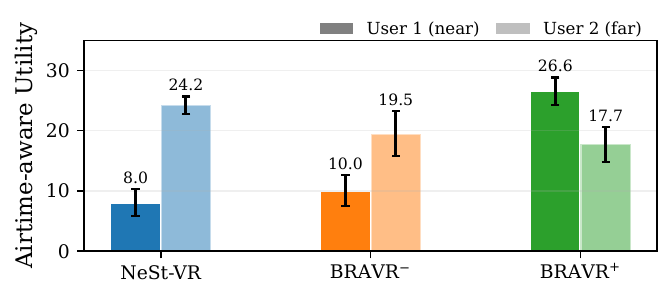}
    \end{subfigure}
    \caption{\textbf{Per-user airtime-aware utility (near--far scenario).} Mean $\pm$ standard deviation across evaluation sessions.}
    \label{fig:utility_mu_far}
\end{figure}

\subsection{Multi-User Scenario}\label{sec:results_multi_user}

Two users run independent sessions using the same bitrate adaptation strategy over a shared channel. Unlike the single-user case, each user's decisions affect the other's adaptation through shared-medium contention.
We test (i)~\emph{near--near}, with both users at Location~A under similarly favorable channel conditions, and (ii)~\emph{near--far}, with User~1 at A and User~2 at B under asymmetric channel conditions.

\subsubsection{Both Users Near the AP (near--near)}
Table~\ref{tab:multi_user_close} reports average per-user performance for all adaptive methods and a CBR baseline at $80$~Mbps per user (higher symmetric rates caused disconnections). Fig.~\ref{fig:utility_mu_close} shows airtime-aware utility.

\vspace{0.08cm}
\noindent\textit{Performance.}
At $80$~Mbps per user, CBR drives the network near saturation, causing sustained latency and packet/frame loss for both users, whereas adaptive methods maintain acceptable mean QoS and achieve comparable rates. In particular, \textsc{BRAVR}$^{+}$ selects slightly lower bitrates but delivers more reliable performance, with a QoS satisfaction above $90\%$. This represents relative gains of $+13.2\%$ over \textsc{NeSt-VR} and $+19.7\%$ over \textsc{BRAVR}$^{-}$, at bitrate reductions of $-7.8\%$ and $-8.2\%$, respectively. Despite occasional transient airtime overuse, it keeps average per-user airtime below the nominal fair-share level (approximately $45\%$ in our setup), indicating effective airtime control. Together with its higher QoS satisfaction, its compliance with the nominal airtime share contributes to achieving the highest airtime-aware utility for both users.

\vspace{0.08cm}
\noindent\textit{Bitrate adaptation.}
Learning-based approaches switch more often and produce less stable trajectories. In particular, \textsc{BRAVR}$^{+}$ switches at $57.5$--$63.2$ changes/min versus $21.6$--$24.5$ for \textsc{NeSt-VR} (Table~\ref{tab:multi_user_close}). In decentralized and uncoordinated settings, this can lead to oscillatory interactions, as bitrate increases from one user reduce the capacity available to the other, triggering compensatory adaptations.

\vspace{0.08cm}
\noindent\textit{Policy behavior.}
Using the same permutation-importance analysis, QoS signals dominate \textsc{BRAVR}$^{+}$ decisions for both users: NFR deviation from target (0.31 and 0.48 for Users~1 and~2, respectively), VF-RTT short-term trend (0.24 and 0.26), and VF-RTT deviation (0.08). AP- and medium-related features, such as stream airtime fraction (0.02) and channel utilization (0.02), show low individual importance. However, these low values do not rule out their role in policy decisions or the reward function.

\vspace{0.08cm}
\noindent\textit{Takeaway:}
\textsc{BRAVR}$^{+}$ maintains high QoS and regulates long-term resource use, whereas overly high fixed rates can exceed capacity and persistently degrade QoS.

\subsubsection{One User Near, One User Far (near--far)}
Due to path loss, the far user (User~2) experiences lower PHY rates under the AP's rate adaptation, increasing airtime per delivered bit and reducing medium efficiency; this can limit the near user's (User~1) performance~\cite{michaelides2025lessons}, motivating bitrate adaptation that accounts for both QoS and shared-medium constraints. Table~\ref{tab:multi_user_far} reports average per-user performance for all adaptive methods and a CBR baseline at $70$~Mbps per user (higher symmetric rates caused frequent disconnections), while Fig.~\ref{fig:utility_mu_far} shows the airtime-aware utility.

\vspace{0.08cm}
\noindent\textit{Performance.}
CBR preserves acceptable QoS for the near user but severely degrades the far user, whereas adaptive methods maintain acceptable mean QoS for both. \textsc{NeSt-VR} and \textsc{BRAVR}$^{-}$ exploit available capacity, achieving higher near-user bitrates but persistently exceeding the near user's nominal airtime share. This increases medium contention, contributing to transient QoS degradations despite acceptable mean performance. In contrast, \textsc{BRAVR}$^{+}$ reduces the near-user bitrate to limit airtime overuse, reducing contention and stabilizing the medium. In particular, it improves QoS satisfaction by $+16.3\%$ and $+23.1\%$ over \textsc{NeSt-VR} and \textsc{BRAVR}$^{-}$ for the near user, and by $+6.8\%$ and $+20.5\%$, respectively, for the far user. These QoS gains come at the cost of lower bitrates for the near user. Overall, \textsc{BRAVR}$^{+}$ achieves its design objective of controlling airtime overuse while maintaining robust QoS, attaining the highest aggregate airtime-aware utility, primarily through gains for the near user. However, the released airtime does not translate into a higher far-user bitrate, likely because its weaker link remains limiting. This motivates airtime regulation that accounts for users' link conditions and their ability to benefit from additional airtime.

\vspace{0.08cm}
\noindent\textit{Policy behavior.}
Using the same permutation-importance analysis, current bitrate dominates the near user's decisions (0.67), as it reflects medium occupancy. Latency-related features, including VF-RTT deviation from target (0.27) and VF-RTT short-term trend (0.06), together with airtime fairness (0.11), also contribute, whereas stream airtime fraction (0.03) and channel utilization (0.02) have minor influence. For the far user, QoS-related features dominate under channel-limited operation. Latency-related features, including peak-to-mean VF-RTT (0.82), VF-RTT short-term trend (0.49), and VF-RTT deviation from target (0.05), are most influential, followed by the reliability-related NFR deviation from target (0.15); bitrate- and medium-related features contribute less.

\vspace{0.08cm}
\noindent\textit{Takeaway:}
\textsc{BRAVR}$^{+}$ maintains QoS for both users while regulating near-user airtime and preventing persistent medium overuse. Under adverse conditions, however, reducing airtime on a strong link does not necessarily improve the weak link's performance.

\section{Conclusions}\label{sec:conclusions}

This paper presents \textsc{BRAVR}, a decentralized AP-assisted reinforcement-learning approach for real-time VR bitrate adaptation over Wi-Fi. It combines application- and AP-level information to balance visual quality, latency, reliability, and multi-user airtime fairness. Experiments on a physical Wi-Fi~5 testbed show that \textsc{BRAVR}$^{+}$ maintains robust QoS while preventing sustained airtime overuse. The results demonstrate improvements over the AP-blind ablation and performance comparable to or better than that of the heuristic \textsc{NeSt-VR}, highlighting the practical value of \textsc{BRAVR}'s AP-assisted adaptation for VR streaming over Wi-Fi. 

Future work will consider more demanding scenarios, including higher user densities and overlapping BSSs, as well as newer Wi-Fi technologies. Further directions include refining airtime-aware adaptation to account for heterogeneous link conditions, exploring multi-objective reinforcement learning to reduce manual reward weighting, and investigating coordinated or centralized control, with the AP serving as a decision or information hub, to improve network-wide efficiency.


\section{Acknowledgements}
We thank Hugo Sanz Benito for their valuable support in conducting the multi-user tests.

This work was supported by the following projects: MLDR (Chist-ERA WAI 2022) PCI2023-145958-2/MCIU/AEI/10.13039/501100011033/UE; REALM (GA 101298050, funded by the EU); TRUE Wi-Fi PID2024-155470NB-I00/MICIU/AEI/10.13039/501100011033/FEDER,UE; ICREA Academia 2024 program (2024 ICREA 00077/AGAUR/Department of Research and Universities of the Government of Catalonia); and MdM CEX2021-001195-M/MICIU/AEI/10.13039/501100011033. The views and opinions expressed are those of the authors and do not necessarily reflect those of the European Union; neither the European Union nor the granting authority can be held responsible for them.

\vspace{0.5cm}

\noindent © IFIP, 2026. This is the author's version of the work. It is posted here by permission of IFIP for your personal use. Not for redistribution. The definitive version was published in IFIP WMNC 2026.


\bibliographystyle{elsarticle-num} 
\bibliography{References}

\end{document}